\documentclass[11pt,a4paper]{article}

\usepackage{epsfig}
\usepackage{graphicx}
\usepackage{amssymb}
\usepackage{amsmath}
\begin{document}

\title{Braneworld black holes as gravitational lenses}

\author{Ernesto F. Eiroa\thanks{e-mail: eiroa@iafe.uba.ar} \\
{\small  Instituto de Astronom\'{\i}a y F\'{\i}sica del Espacio, C.C.
67, Suc. 28, 1428, Buenos Aires, Argentina}}
\maketitle

\begin{abstract}
Black holes acting as gravitational lenses produce, besides the primary and 
secondary weak field images, two infinite sets of relativistic images. These 
images can be studied using the strong field limit, an analytic method based 
on a logarithmic asymptotic approximation of the deflection angle.  
In this work, braneworld black holes are analyzed as gravitational lenses 
in the strong field limit and the feasibility of observation of the 
images is discussed.
\end{abstract}



\section{Introduction}\label{intro}

Braneworld cosmologies \cite{maartens} have attracted great attention from 
researchers in the last few years. In these cosmological models, the ordinary 
matter is confined to a three dimensional space called the brane, embedded in 
a larger space called the bulk in which only gravity can propagate. 
Cosmologies with extra dimensions were proposed in order to solve the 
hierarchy problem, that is to explain why the gravity scale is sixteen 
orders of magnitude greater than the electro-weak scale, and are
motivated by recent developments of string theory, known as M-theory. 
The study of black holes on the brane is rather 
difficult because of the confinement of matter on the brane whereas the 
gravitational field can access to the bulk. The full five dimensional bulk 
field equations have no known exact solutions representing static and 
spherically symmetric black holes with horizon on the brane. Instead, 
several braneworld black hole solutions have been found based on different 
projections on the brane of the five dimensional Weyl tensor.\\ \indent
Since the publication of the paper of Virbhadra and Ellis \cite{virbha} there 
has been a growing interest in the study of lensing by black holes. They 
analyzed numerically a Schwarzschild black hole at the Galactic center 
acting as a gravitational lens.  For black hole gravitational lenses, large 
deflection angles are possible for photons passing close to the photon 
sphere. These photons could even make one or more complete turns, in both 
directions of rotation, around the black hole before eventually reaching an 
observer. As a consequence, two infinite sequences of images, called 
relativistic images, are formed at each side of the black hole. Instead of 
making a full numerical treatment, a logarithmic approximation of the 
deflection angle can be done to obtain the relativistic images. This 
approximation was first used by Darwin \cite{darwin} for Schwarzschild black 
holes, rediscovered and called the strong field limit by Bozza \textit{et al.} 
\cite{bozcap}, extended to Reissner--Nordstr\"{o}m geometries by Eiroa 
\textit{et al.} \cite{eiroto}, and generalized to any spherically symmetric 
black hole by Bozza \cite{bozza}. The strong field limit was subsequently 
applied to retrolensing by Eiroa and Torres \cite{eitor} and used by other 
authors in the analysis of different lensing scenarios. The study of 
gravitational lensing by braneworld black holes could be useful in the 
context of searching possible observational signatures of these objects.

\section{The strong field limit}\label{sflim}

\begin{figure}[h]
\vspace{-0.5cm}
\begin{center}
\includegraphics[width=12cm]{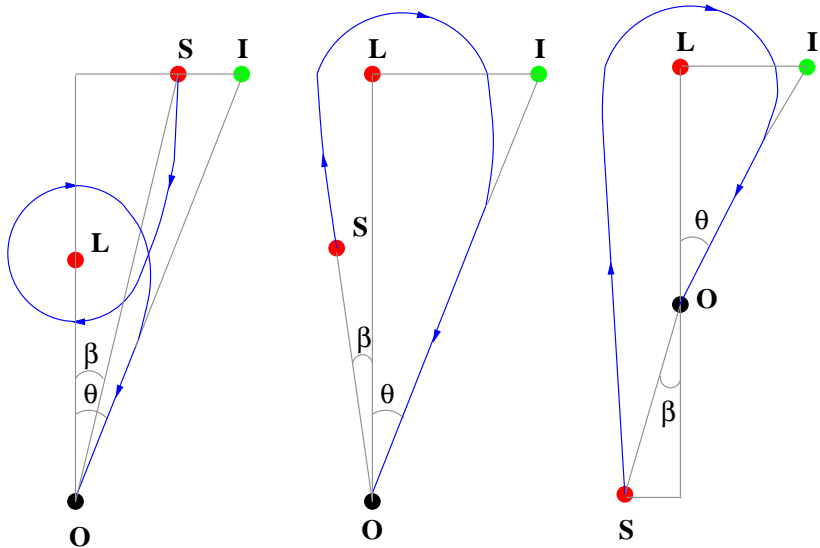}
\end{center}
\vspace{-0.5cm}
\caption{Schematic diagrams of possible lensing geometries.}
\label{fig}
\end{figure}

In Fig. \ref{fig}, three possible lensing situations are shown 
schematically. In the left, the black hole, which will 
be called the lens (l), is between the source (s) and the observer (o); in the 
middle, the source is between the lens and the observer; and in the right the 
observer is between the source and the lens. The first case will be called 
standard lensing and the other two, respectively, cases I and II of 
retrolensing.
The observer-source, observer-lens and the lens-source distances, here taken 
much greater than the horizon radius $r_{h}$, are, in units of $r_{h}$,  
$d_{os}$, $d_{ol}$ and $d_{ls}$, respectively. Defining $\beta $ as 
the angular position of the point source and $\theta $ as the 
angular position of the images (i), both seen from the observer, 
and $\alpha $ as the deflection angle of the photons, the lens 
equation \cite{virbha,eitor} has the form:
\begin{equation}
\tan \beta =\tan \theta -c_{3}\left[ \tan (\alpha -\theta)
+\tan \theta \right] ,
\label{e1}
\end{equation}
where $c_{3}=d_{ls}/d_{os}$ for standard lensing and 
$c_{3}=d_{os}/d_{ol}$ or $c_{3}=d_{os}/d_{ls}$ for 
cases I and II of retrolensing, respectively. $\beta $ can be taken positive
without losing generality. For a spherically symmetric black hole with 
asymptotically flat metric:
\begin{equation}
ds^{2}=-f(x)dt^{2}+g(x)dx^{2}+h(x)d\Omega ^{2},
\label{e2}
\end{equation}
where $x=r/r_{h}$ is the radial coordinate in units of the horizon radius, 
the deflection angle $\alpha $ as a function of the closest approach 
distance $x_{0}$ is given by \cite{weinberg}
\begin{equation}
\alpha (x_{0})=-\pi +\int_{x_{0}}^{\infty }
2\left[ \frac {g(x)}{h(x)}\right] ^{1/2}
\left[ \frac {h(x)f(x_{0})}{h(x_{0})f(x)}-1\right] ^{-1/2}dx.
\label{e3}
\end{equation}
There are two cases where the deflection angle can be approximated by simple 
expressions:
\begin{itemize}
\item Weak field limit: for $x_{0}\gg x_{ps}>1$ 
($x_{ps}$ is the photon sphere radius), a first non null order  
Taylor expansion in $1/x_{0}$ \cite{schneider} is made.
\item Strong field limit: $\alpha (x_{0})$ diverges when 
$x_{0}= x_{ps}$, and for $0<x_{0}-x_{ps}\ll 1$, it can be approximated 
by a logarithmic function \cite{bozza}:
\begin{equation}
\alpha(x_{0})\approx -a_{1}\ln(x_{0}-x_{ps})+a_{2},
\label{sfl}
\end{equation}
where $a_{1}$ and $a_{2}$ are constants. 
\end{itemize}
The weak field limit is used for small deflection angles, as it happens when 
the lens is a star, a galaxy, or for the weak field images produced by photons 
with large impact parameters in the case of black holes. The strong field 
limit is useful to make an approximate analytical treatment of the 
relativistic images for black hole lenses. \\

The closest approach distance $x_{0}$ is related with the impact parameter $b$ 
(in units of the horizon radius) by the equation $b=h(x_{0})/f(x_{0})$ 
\cite{weinberg}, and from the lens geometry $b=d_{ol}\sin \theta $, 
so $x_{0}$ can be calculated as a function of $\theta $. By putting 
$x_{0}(\theta )$ in Eq. (\ref{sfl}) to have $\alpha (\theta )$, then 
replacing $\alpha (\theta )$ in Eq. (\ref{e1}), and finally inverting the lens 
equation (\ref{e1}), the positions of the images are obtained as a 
function of $\beta $ and the distances involved.

\section{Five dimensional Schwarzschild black hole lens}\label{5dschw}

In this Section, a five dimensional Schwarzschild black hole is studied as 
gravitational lens in the context of braneworlds. The cosmological model 
adopted is Randall-Sundrum type II \cite{rs}, which consist of a positive 
tension brane in a one extra-dimensional bulk with negative cosmological 
constant. For this black hole, the four dimensional induced metric on the 
brane is:
\begin{equation}
ds^{2}=-\left( 1-\frac{r_{h}^{2}}{r^{2}}\right)dt^{2}+
\left( 1-\frac{r_{h}^{2}}{r^{2}}\right)^{-1}dr^{2}+r^{2}d\Omega ^{2},
\label{e4a}
\end{equation}
where $d\Omega ^{2}= d\vartheta ^{2}+\sin^{2}\vartheta d\varphi ^{2}$ and
\begin{equation}
r_{h}=\sqrt{\frac{8}{3\pi }}\left( \frac{l}{l_{4}}\right)^{1/2} 
\left( \frac{M}{M_{4}}\right)^{1/2}l_{4},
\label{e4b}
\end{equation}
with $l<0.1$ mm \cite{long} the AdS radius, $l_{4}$ and $M_{4}$, respectively, 
the Planck length and mass (units in which $c=\hbar =1$ are used). The 
main features of these braneworld black holes are \cite{kanti}:
\begin{itemize}
\item If $r_{h}\ll l$ they are a good approximation, near the event horizon, 
of black holes produced by collapse of matter on the brane.
\item Primordial black holes in this model have a lower evaporation rate by
Hawking radiation than their four dimensional counterparts in standard 
cosmology, and they could have survived up to present times.
\item Only energies of about $1$ TeV are needed to produce black holes 
by particle collisions instead of energy scales about $10^{16}$ TeV 
required if no extra dimensions are present. These small size black holes 
could be created in the next generation particle accelerators or 
detected in cosmic rays.
\end{itemize}
Majumdar and Mukherjee \cite{majumuk} considered gravitational lensing 
in the weak field limit for the black holes discussed above. The positions and 
magnifications of the relativistic images in the strong field limit are 
obtained bellow (for more details, see the paper by Eiroa \cite{eiroa}). 
  
\subsection{Deflection angle}

For the braneworld black hole metric (\ref{e4a}) the deflection angle is
\begin{equation}
\alpha (x_{0})=-\pi +2x_{0}^{2}\int_{x_{0}}^{\infty }
\left[ x^{4}(x_{0}^{2}-1)-x_{0}^{4}(x^{2}-1)\right] ^{-1/2}dx,
\label{e5}
\end{equation}
which, near the photon sphere ($x_{ps}=\sqrt{2}$), can be approximated by 
\cite{eiroa}
\begin{equation}
\alpha (x_{0})=-\sqrt{2}\ln(x_{0}-\sqrt{2})+\sqrt{2}\ln 4-\pi 
+O(x_{0}-\sqrt{2}).
\label{e6}
\end{equation}
Using the impact parameter $b=x_{0}^{2}/\sqrt{x_{0}^{2}-1}$, it takes the form
\begin{equation}
\alpha (b)=-\frac{\sqrt{2}}{2}\ln \left( \frac{b}{b_{ps}}-1 \right) 
+\sqrt{2}\ln(4\sqrt{2})-\pi +O(b-b_{ps}). 
\label{e7}
\end{equation} 

\subsection{Image positions}

In case of high alignment, $\beta \ll 1$, $\theta \ll 1$ and $\alpha $ takes 
values close to multiples of $\pi $. There are two sets of relativistic images.
For the first one, $\alpha =m\pi +\Delta \alpha _{m}$, with 
$0<\Delta \alpha _{m}\ll 1$, $m=2n$ for standard lensing and $m=2n-1$ 
for retrolensing ($n\in \mathbb{N}$). The other set of images  
have $\alpha =-m\pi -\Delta \alpha _{m}$. Then, the lens equation can be 
approximated by
\begin{equation}
\beta =\theta \mp c_{3}\Delta \alpha _{m},
\label{e8}
\end{equation}
and from the geometry of the system
$b=d_{ol}\sin \theta \approx d_{ol}\theta $,
so, defining $\theta _{ps}=2/d_{ol}$, the deflection angle is given by
\begin{equation}
\alpha (\theta )=-c_{1}\ln \left( \frac{\theta}{\theta _{ps}}-1 \right) 
+c_{2}+O(\theta-\theta _{ps}), 
\label{e9}
\end{equation}
where $c_{1}=\sqrt{2}/2$ and $c_{2}=\sqrt{2}\ln(4\sqrt{2})-\pi $.
Inverting Eq. (\ref{e9}), making a first order Taylor expansion
around $\alpha =\pm m\pi $, and using Eq. (\ref{e8}), the angular position of 
the $m$-th image is:
\begin{equation}
\theta _{m}=\pm \theta ^{0}_{m}+\frac {\zeta _{m}}{c_{3}}(\beta 
\mp \theta ^{0}_{m}),
\label{e10}
\end{equation}
with
\begin{equation}
\theta ^{0}_{m}=\theta _{ps}\left[ 1+e^{(c_{2}-m\pi )/c_{1}}
 \right],
\label{e11}
\end{equation}
and
\begin{equation}
\zeta _{m}=\frac{\theta _{ps}}{c_{1}}e^{(c_{2}-m\pi )/c_{1}}.
\label{e12}
\end{equation}
For $\beta =0$ (perfect alignment), instead of point images, an infinite 
sequence of Einstein rings with angular radii
\begin{equation}
\theta ^{E}_{m}=\left( 1-\frac {\zeta _{m}}{c_{3}}\right) \theta ^{0}_{m},
\label{e13}
\end{equation}
is obtained.

\subsection{Image magnifications}

As gravitational lensing conserves surface brightness \cite{schneider}, the 
magnification of the $m$-th image is the quotient of the solid angles 
subtended by the image and the source:
\begin{equation}
\mu _{m}=\left| \frac{\sin \beta }{\sin \theta _{m}}
\frac{d\beta }{d\theta _{m}}\right|^{-1}\approx 
\left| \frac{\beta }{\theta _{m}} \frac{d\beta }{d\theta _{m}}\right|^{-1};
\label{mu1}
\end{equation}
so, using Eq. (\ref{e10}) and keeping only the first order term in 
$\zeta _{n}/c_{3}$, it is easy to see that
\begin{equation}
\mu _{m}=\frac{1}{\beta}\frac{\theta ^{0}_{m}\zeta _{m}}{c_{3}},
\label{mu2}
\end{equation}
for both sets of images. The total magnification $\mu $ is found by 
summing up the magnifications of all images; then, for standard lensing 
$\mu $ is given by
\begin{equation}
\mu =\frac {8}{\beta }
\frac{e^{c_{2}/c_{1}}\left( 1+e^{c_{2}/c_{1}}+e^{2\pi /c_{1}}\right) }
{d_{ol}^{2}c_{1}c_{3}(e^{4\pi /c_{1}}-1)},
\label{mu4}
\end{equation}
and for retrolensing by 
\begin{equation}
\mu =\frac {8}{\beta }
\frac{e^{(c_{2}+\pi )/c_{1}}\left[ 1+e^{(c_{2}+\pi )/c_{1}}+e^{2\pi /c_{1}}
\right] }{d_{ol}^{2}c_{1}c_{3}(e^{4\pi /c_{1}}-1)}.
\label{mu5}
\end{equation}
When $\beta =0$ the point source approximation fails because the 
magnifications diverge. Then, an extended source analysis is needed. In this 
case, it is necessary to integrate over its luminosity profile to obtain the 
magnification of the images. If the source is an uniform disk 
$D(\beta _{c},\beta _{s})$, with angular radius $\beta _{s}$ and centered in 
$\beta _{c}$ (taken positive), the magnification of the $m$-th image is
\begin{equation}
\mu _{m}=\frac{I}{\pi \beta _{s}^{2}}\frac {\theta ^{0}_{m}\zeta _{m}}{c_{3}},
\label{ext3}
\end{equation}
where
\begin{equation}
I=2[(\beta _{s}+\beta _{c})E(k)+(\beta _{s}-\beta _{c})K(k)] ,
\label{ext6}
\end{equation}
with $K(k)=\int_{0}^{\pi /2}\left( 1-k^2\sin ^{2}\phi \right) ^{-1/2}d\phi $ 
and $E(k)=\int_{0}^{\pi /2}\left( 1-k^2\sin ^{2}\phi \right) ^{1/2}d\phi $, 
respectively, the complete elliptic integrals of first and second kind with 
argument $k=2\sqrt{\beta _{s}\beta _{c}}/(\beta _{s}+\beta _{c})$. Then, the 
total magnification of an uniform source for standard lensing is
\begin{equation}
\mu =\frac {8I}{\pi \beta _{s}^{2}}
\frac{e^{c_{2}/c_{1}}\left( 1+e^{c_{2}/c_{1}}+e^{2\pi /c_{1}}\right) }
{d_{ol}^{2}c_{1}c_{3}(e^{4\pi /c_{1}}-1)},
\label{ext4}
\end{equation}
and for retrolensing
\begin{equation}
\mu =\frac {8I}{\pi \beta _{s}^{2}}
\frac{e^{(c_{2}+\pi )/c_{1}}\left[ 1+e^{(c_{2}+\pi )/c_{1}}+e^{2\pi /c_{1}}
\right] }{d_{ol}^{2}c_{1}c_{3}(e^{4\pi /c_{1}}-1)}.
\label{ext5}
\end{equation}
These expressions always give finite magnifications.

\section{Other braneworld black hole lenses}\label{other}

Whisker \cite{whisker}, using the Randall--Sundrum II cosmological model, made 
a strong field limit analysis for two possible braneworld black hole 
geometries. The tidal Reissner-Nordstr\"om black hole \cite{dadhich} has the 
metric on the brane:
\begin{equation}
ds^{2}=-\left( 1-\frac{2GM}{r} +\frac{Q}{r^{2}} \right) dt^{2}+
\left( 1-\frac{2GM}{r} +\frac{Q}{r^{2}} \right) ^{-1}dr^{2}+r^{2}
d\Omega ^{2},
\label{w1}
\end{equation}
where the tidal charge parameter $Q$ comes from the projection on the brane of 
free gravitational field effects in the bulk, and it can be positive or 
negative. When $Q$ is positive, it weakens the gravitational field, and if it 
is negative the bulk effects strengthen the gravitational field, which is 
physically more natural. This metric has the same properties as the 
Reissner-Nordstr\"om geometry for $Q>0$, there are two horizons, both of which 
lie within the Schwarzschild horizon. When $Q<0$, there is one horizon, lying 
outside Schwarzschild horizon. For all $Q$, there is a singularity at 
$r=0$. The event horizon radius is given by  
$r_{h}=r_{+}=GM+\sqrt{(GM)^{2}-Q}$ and the radius of the photon sphere by
\begin{equation} 
r_{ps} =\frac{3}{2}GM+\frac{1}{2}\sqrt{9(GM)^{2}-8Q}.\nonumber 
\end{equation}
The strong field limit coefficients $c_{1}$ and $c_{2}$ has to be obtained 
numerically and they are plotted in Fig. 4 of Ref. \cite{whisker}. \\

Whisker \cite{whisker} also proposed as a \textit{working metric} for the 
near-horizon geometry the ``$U=0$'' solution: 
\begin{equation}
ds^2 = -\frac{(r-r_h)^2}{(r+r_t)^2}dt^2 +
\frac{(r+r_t)^4}{r^4}dr^2+\frac{(r+r_t)^4}{r^2}d\Omega ^2, 
\label{w2}
\end{equation}
which has the horizon at $r=r_h$. This metric has several differences to 
the standard Schwarzschild geometry; the horizon is singular (except when 
$r_h=r_t$, for which is just the standard Schwarzschild solution in 
isotropic coordinates), and the area function has a turning point at 
$r=r_t$, that can be either inside or outside the horizon. Another difference, 
coming from that $g_{tt} \ne g_{rr}^{-1}$, is that the ADM mass and 
the gravitational mass (defined by $g_{tt}$) are not the same. 
The photon sphere has radius:
\begin{equation}
r_{ps}=r_{h}+r_{t}+\sqrt{r_{h}^{2}+r_{h}r_{t}+r_{t}^{2}}.\nonumber
\end{equation}
The coefficients $c_{1}$ and $c_{2}$ for the metric ``U=0'' are plotted in 
Fig. 3 of Ref. \cite{whisker}.\\

The expressions for the positions and magnifications of the relativistic 
images obtained in Sec. \ref{5dschw} are valid for these metrics, replacing  
$c_{1}$ and $c_{2}$ by the values obtained by Whisker, and using the 
corresponding value of $r_{h}$ to adimensionalize the distances. The 
metrics considered in this Section, unlike those studied in Sec. \ref{5dschw}, 
can be applied to massive astrophysical black holes.
In Ref. \cite{whisker}, the case of the supermassive black hole in 
the Galactic center was analyzed and the results for the two geometries 
described above were compared with those corresponding to the standard four 
dimensional Schwarzschild black hole.\\

Another related work, using the Arkani-Hamed, Dimopoulos and Dvali braneworld 
model, is that by Frolov \textit{et al.} \cite{frolov}. They found 
the induced metric on the brane by a Schwarzschild black hole moving in 
the bulk, and also studied the deflection of light on the brane produced by 
the black hole.

\section{Final remarks}\label{conclu}

The relativistic images produced by black hole lenses have an angular 
position about the size of the angle subtended by the photon sphere and they 
are strongly demagnified, making their observation extremely difficult. In 
astrophysical scenarios, the observation of these relativistic images is 
not possible today and it will be a challenge for the next decade. 
Whisker \cite{whisker} have shown that a braneworld black hole in the center 
of our galaxy could have different observational signatures than the four 
dimensional Schwarzschild one. In the case of small size black holes studied 
as gravitational lenses by Eiroa \cite{eiroa}, the observation of the 
relativistic images in astrophysical contexts will be even more difficult. 
But if the braneworld model is correct and these black holes can be produced 
by next generation of particle accelerators or in cosmic ray showers, it 
opens up the possibility of observing the phenomena of lensing by black holes 
in the laboratory.

\section*{Acknowledgments}
The author is grateful to the organizers of the conference \textit{100 Years 
of Relativity} for financial assistance. This work has been partially 
supported by UBA (UBACYT X-103).

\end{document}